# Explaining quanta with optical illusions

*Gianluca Li Causi*

*gianluca.licausi@inaf.it*

**INTRODUCTION / ABSTRACT**

In this work, I propose a way to help high school students and the general population understand quantum concepts by adopting a new inherently dual representation. Major difficulties in explaining to people the basic concepts of quantum mechanics reside in the apparent impossibility of representing quantum superposition with examples taken from everyday life. In this context, I propose a new pictorial paradigm that illustrates a number of quantum concepts by means of optical illusions, potentially without raising misconceptions. The method is based on "bistable reversible figures," which induce in the viewer a multistable perception, conveying a direct understanding of superposition, random collapse, and observer effect via a sensorial experience. I present the advantages and discuss the limitations of this analogy, and show how it extends to the concepts of complementarity and quantum entanglement, also helping to avoiding misconceptions in quantum teleportation. Finally, I also address quantum spin and quantum measurement by using different types of optical illusions.

**QUANTUM STATES AND SUPERPOSITION**

Quantum mechanics demonstrated that an elementary particle has qualities, called "quantum properties", which can only take discrete values upon measurement. An electron, for example, possesses a property called "spin" that, when measured, can only be found in one of two "quantum states", commonly called "up" and "down", and formally indicated as $|+\rangle$ and $|-\rangle$, or $|0\rangle$ and $|1\rangle$ in the qubit notation that I adopt hereafter. Other particles can show more than two exclusive states after measurements of their properties. The set of states assumed by a particle for its various properties defines its overall "quantum state".

A classical object too has a "state", which is the collection of its qualities, like the color, the size, the weight, and so on. However, a classical object can only exist in one state at a time, e.g. it cannot be simultaneously yellow and blue, while experiments clearly show that a quantum object can exist in a "superposition state", which means that it can be at the same time in more than one state. Specifically, the "superposition principle" of quantum mechanics says that one can express any quantum state as the superposition of other quantum states. Although, when someone "observes" a particle in superposition, i.e. take a measurement of a quantum property, they see only one of the possible discrete outcomes, which value being completely random and inherently unpredictable.

The fact the classical objects cannot behave in this way makes it very difficult for people to realize what the "quantum object" actually is, because they need to imagine something having exclusive qualities at the same time. To help with this, scientists proposed cute explanation methods, which use diagrams, colors, games, simulations, or everyday objects[1,2,3,4,5]. Nonetheless, the difficulty remains, largely due to the *lack of a representation that is inherently dual*. This is why I propose here to think of a quantum object as a "bi-stable drawing".

**THE BI-STABLE DRAWING REPRESENTATION**

A bi-stable drawing is a picture that creates ambiguity between two distinct shapes in the interpretation by human visual system. Looking at such images causes a spontaneous and stochastic alternation between two mutually exclusive perceptual states[6].

Such pictures are also called "reversible optical illusions", the most famous of which are the "Necker cube" [7], the "My wife and my mother-in-law" [8], and the "Rubin vase" [9], but there are a lot.

The first of them (Fig. 1) is the 2D parallel projection of a wireframe cube, which can be interpreted to have either the lower-left or the upper-right square as its front side. In the "My wife and my mother-in-law", (Fig. 2) the profile of a young girl appears when the face of an old woman disappears. Finally, in the "Rubin vase" (Fig. 3) that belongs to the category of "figure-ground illusions", one can perceive a white vase on a dark background or, alternatively, the dark profile of two faces looking to each other onto a white background.

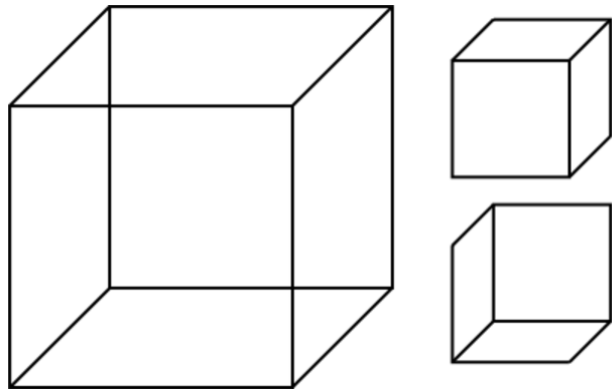

*Fig. 1: The "Necker cube" (left) can be perceived as a cube seen from above or from behind (right).*

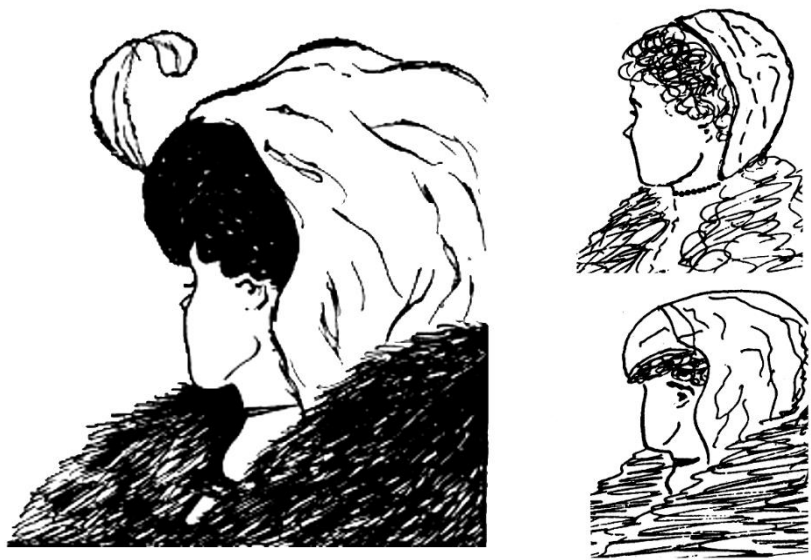

*Fig. 2: The "My wife and my mother-in-law" drawing (left) can be seen as a girl or an old woman (right).*

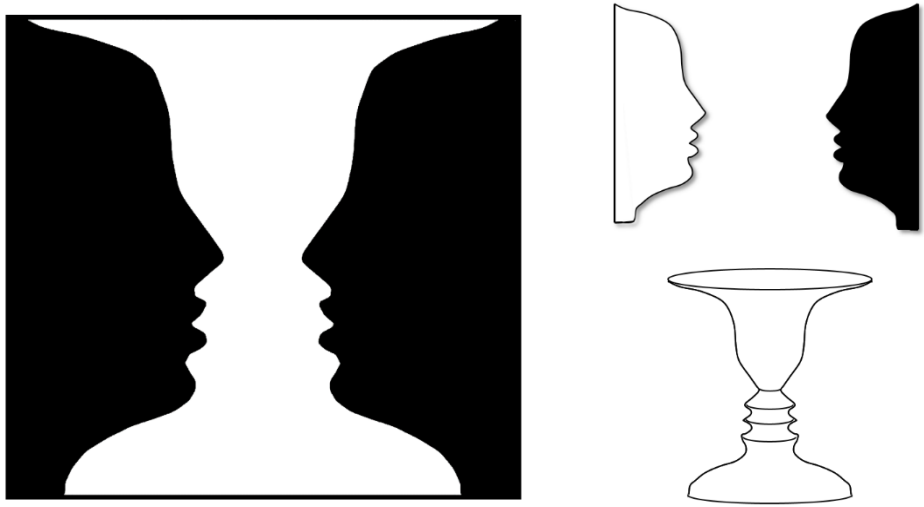

*Fig. 3: In the "Rubin vase" drawing (left), one can see two faces or one vase (right).*

The peculiarity of such drawings is that, every time one looks at them, he perceives a shape or its complement in a purely random way. Therefore, a bi-stable drawing behaves like a quantum object, which is dual and, when observed, instantly "collapses" into one of its exclusive states in a completely random way.

Some psychological studies suggested such similarity between human perception and quantum mechanics with the scope to derive a quantum-like model for brain mental states[10,11,12]. On the contrary, I propose here to adopt this similarity to explain quantum concepts easily.

In fact, this apparently simple analogy includes a proper correspondence for a number of physical aspects of the quantum theory, namely:

- *quantum object* → bi-stable drawing
- *pure quantum states |0> and |1>* → each one of the two shapes
- *quantum superposition* → coexistence of the two shapes
- *quantum collapse* → sudden perception of a single shape
- *measurement* → the act of looking at the drawing
- *random result of a measurement* → randomness in which shape is perceived
- *orthogonality of the pure states* → impossibility to perceive the two shapes simultaneously
- *quantum de-coherence* → loss of stable perception of one shape
- *a quantum is neither a wave or a particle, it's just a different thing* → a drawing is neither of its perceived shapes, it's just a distribution of black and white regions so that bi-stable drawings offer a new possibility to the spectator to realize in a personal way these very concepts.

Note that this game does not allow cutting out a single shape from the drawing, because the isolated shape is no more a bi-stable drawing and so cannot represent a quantum object.

The last item in the list deserves some more words. The viewer is led to think of these drawings as being in a state of "suspended reality" that becomes real in a different way only after observation. However, one cannot ask whether the drawing of the Rubin vase depicts "in reality" a vase or two faces, or what is "in reality" the front face of the Necker cube. In fact, *a drawing is not what it represents, but only a distribution of black and white regions, and only the interaction between the drawing and the observer generates a meaning*. This concept helps to understand the wave-particle duality, suggesting how mutually exclusive properties like wave and particle can be aspects of something inherently different, the quantum object, exhibiting either wave or particle behavior depending on what is observed.

Of course, my paradigm remains an analogy and, like all the analogies, it has some limitations.

The main limitation is that it does not support measurement. In fact, it cannot reproduce the stability of a quantum state after collapse. If one performs the same measurement on a quantum state soon after a previous measurement, the result is always the same; instead, the perception of a bi-stable drawing continues to swap between the two alternatives as long as one looks at it.

The teacher would need to make it clear that repeating measurements would be akin to seeing the old woman and then only seeing the old woman each time the same drawing was sequentially observed. Furthermore, instead of seeing this as a problem, he could leverage on this feature to represent the "quantum de-coherence", a very important aspect in the field of quantum computing; this is why I added it in the previous list.

Finally, a clever spectator knows well that the described perception is just a response of his visual system and brain, and that nothing really changes in a bi-stable drawing when someone looks at it. Such knowledge could lead the viewer to think that also the quantum state of a real particle only resides in the observer and that the quantum collapse is a pure illusion. This is wrong in the common Copenhagen interpretation, which holds that quantum descriptions are objective; however, different interpretations of quantum mechanics exist, for which there is no objective truth independent on the observer. So again, the reader can use this inadequacy of the model to illustrate those different interpretations.

It's worth noting that of course these drawings are only able to represent dual states which have equal probabilities to be measured.

**COMPLEMENTARITY PRINCIPLE**

I further exploit the proposed correspondence to get a representation of the "complementarity principle" [13], which states that quantum objects have certain pairs of complementary properties that cannot be observed simultaneously. For example, measurements of spin components along perpendicular axes are complementary, as the observation of one leaves the others in superposition.

In the proposed paradigm, one can represent a pair of complementary properties with two bi-stable drawings arranged to be seen exclusively.

For example, one could place the "Rubin vase" on the front side of a card and the "My wife and my mother-in-law" on the rear side (Fig. 4). If in the front picture the spectator randomly recognized a vase, and he is now looking one shape in the drawing of the rear side, when turning the card back he will perceive a vase or two faces again randomly, independently on the shape saw before.

In practice, in the moment that the spectator observes one drawing he loses the identity seen in the other, as it happens in a quantum object, where the measurement of a property turns its complement to the undefined state of superposition.

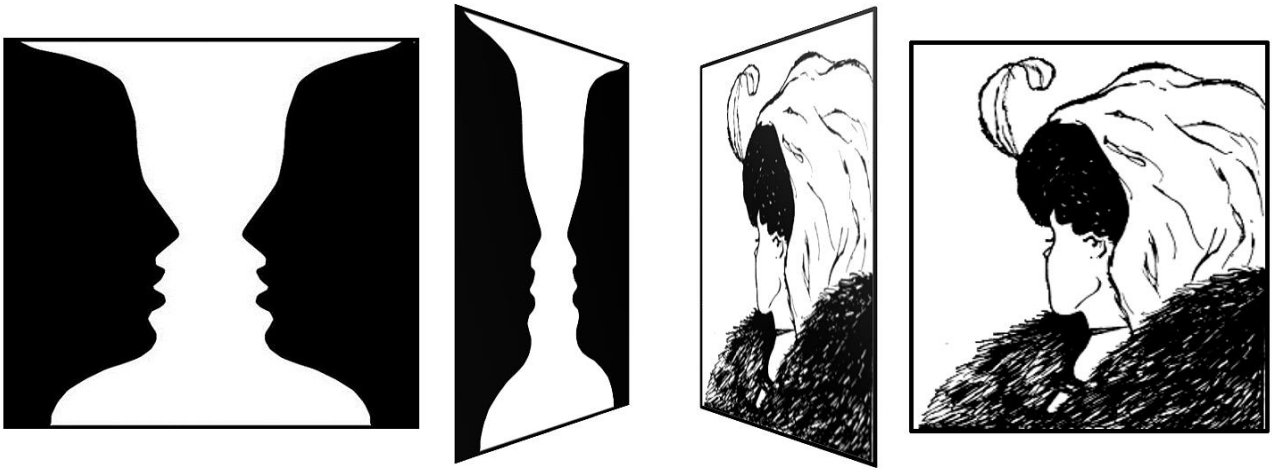

*Fig. 4: A flipping card with a different bi-stable drawing in its front and rear side, can represent the complementary principle.*

The complementarity principle, by excluding the possibility of a simultaneous collapse of two complementary properties, explains why quanta have to be intrinsically dual, and why all particles must have at least one property undefined at a given time, with the consequence that randomness is inherent in the fundamental nature of the universe.

Therefore, I add the following correspondences to the previous list of analogies:

- *pair of complementary properties* → pair of bi-stable drawings in the front and rear side of a card
- *complementarity principle* → loss of perception of a drawing while looking at the other.

As before, it's worth to remind that the teacher shall avoid forcing the analogy to reproduce the measurement process, like e.g. telling that the rear face of the card can be simultaneously viewed (e.g. by placing a mirror), or that seeing a vase and flipping the card 360 degrees to the same illusion doesn't necessarily show the vase again. Such operations shall not be allowed in this game.

**QUANTUM ENTANGLEMENT**

When two quanta interact to each other, a relation establishes between their quantum states and holds until one of them is subjected to a new interaction[14]. Such relation is called "entanglement", or "quantum correlation", and it is at the heart of the disparity between classical and quantum physics.

Consider an experiment in which two particles enter in a region where they interact, and then goes out in different directions.

Let us consider the simple case of a binary property, like the spin of an electron, and suppose that each particle enters the experiment in a superposition state. Before the interaction the two particles are independent, which means that measuring their spins one would find |0> or |1> randomly and independently.

After the interaction, however, the two particles are no longer independent, as laboratory experiments demonstrate. In particular, the spin measured on a particle is always concordant, or always discordant, with the spin measured on the other. In other words, after the interaction the two spins can only be correlated (i.e. |0>, |0> or |1>, |1>), or anti-correlated (i.e. |0>, |1> or |1>, |0>), depending only on how the experiments is conceived. Moreover, such correlation is present even if the time delay between

measurements of first and second spin is less than the time a light ray would need to go from one particle to the other. This excludes any possibility of communication between them and implies quantum non-locality[15].

This behavior rises two questions: how can two things be concordant or discordant if they are still undefined? And, how can the second spin be dependent on the first one if there has been no communication between them?

The presented analogy offers to the spectator a visual answer to both questions.

To visualize the entanglement one places two identical bi-stable drawings side by side, e.g. two copies of the "Rubin vase" (Fig. 5). If looking at the first, one randomly see a vase, after moving the glance to the other, he will see a vase also in it. On the contrary, if somebody sees two faces in the first drawing, they will see two faces in the second too. In other words, in the second drawing one always perceives the same shape randomly perceived in the first, as it happens in a positively correlated entanglement.

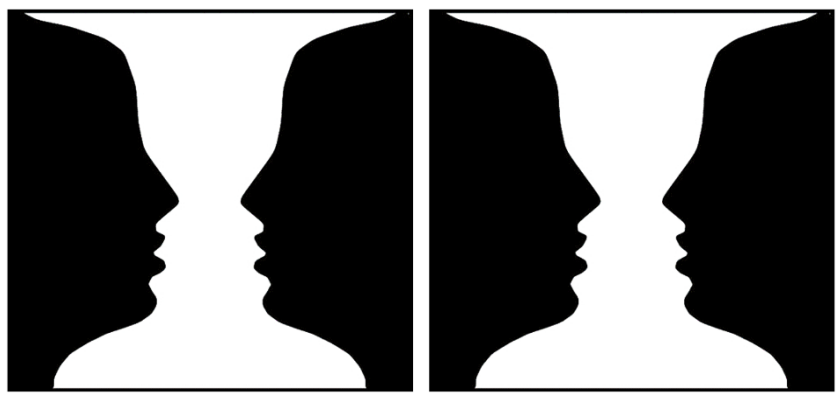

*Fig. 5: Representation of an entanglement in positive correlation.*

Instead, if one aims at illustrating an anti-correlated entanglement, one just have to replace the second drawing with its inverse, where the vase is black and the faces are white (Fig. 6).

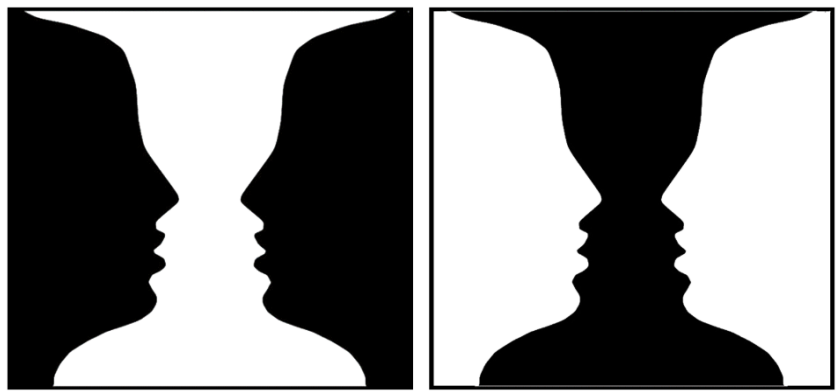

*Fig. 6: Representation of an entanglement in anti-correlation.*

This experiment gives a hint to the spectator on the way two quanta in superposition can be concordant or discordant even if the result of the first measurement is purely random. Moreover, the spectator clearly realizes that the first drawing does not need to communicate anything to the other, because the mutual pattern of black and white regions encodes the type of correlation from the beginning.

In this way, my analogy suggests the correct understanding that quantum correlation is a property encoded in the couple, which has to be considered as a whole. Therefore, one can represent even the strangest of all quantum concepts, i.e. the entanglement, by leveraging the following correspondences:

- *pair of quanta in entanglement* → pair of identical bi-stable drawings placed side-by-side
- *pair of quanta in anti-correlation* → a bi-stable drawing placed aside its inverted copy
- *absence of communication between the quanta* → absence of communication between the drawings

The limitation in this correspondence is that it cannot reproduce the fact that two entangled particles are still correlated if measured by different observers, because there is no correlation in the shapes perceived by two persons that looks at identical bi-stable drawings. Hence, one shall not force the analogy in such direction. In the last section hereafter, I suggest how to overpass this limitation.

**QUANTUM TELEPORTATION**

In quantum physics, teleporting a particle from a place to another means transferring its quantum state to destination without physically moving the particle from departure. This is achieved by exploiting quantum entanglement[16].

To begin, one must prepare a concordant entangled state of two quanta (e.g. two photons) and holds one of them, A, in the departure lab while the other, B, is *physically carried* to destination. This separated couple takes the name of "quantum channel".

What one wants to do is to transfer the quantum state of a third photon, C (called the "teleported photon"), from departure to destination, using the quantum channel as a carrier.

To initiate this transfer one correlates the photon C with the photon A of the quantum channel. Let us call this action the "teleportation entanglement".

Quantum theory tells that, after this correlation, the teleported photon, C, results entangled with the photon B at destination, even if they have never interacted directly. Hence, if C is found e.g. in state |0> also B will be found in state |0> after measurement, which means that the state of a quantum at departure has been transferred into the state of a quantum at destination. Namely, it has been "teleported".

This happens because quantum correlation has a sort of "transitive" property: if one has correlated A with B and then he correlates C with A, then C and B results correlated (while the original correlation between A and B breaks out).

However, all that is true only if the teleportation entanglement between A and C happened concordant. In fact, if it happened discordant the photon B receives the *opposite* state, and people at destination need to *flip* the state of B before measurement if they want to get the same state of C.

The sign of correlation though is random, so that the teleportation can only work if the departure lab informs the destination lab about the sign effectively taken by the correlation. This is a classical information that can only be sent at the speed of light, e.g. via radio, which is the reason why the quantum teleportation cannot be instantaneous and needs the aid of a classical communication channel.

I have not found a way to use three bi-stable drawings to replicate the teleportation protocol and get to perceive in the drawing at destination the same shape perceived in the drawing to be transferred. This probably goes beyond the power of the proposed analogy. Nonetheless, bi-stable drawings turn to be very helpful to show that *quantum teleportation does not actually transport anything*, because the identity of the teleported photon *is already at destination* since we set the quantum channel. In fact, the photon B that we carried to destination was in a state of superposition, so that we *physically carried* all these identities in one move. A spectator who looks at Fig. 5 will immediately understand this fact, because he quickly realizes that physically moving either drawing moves both shapes together.

**SPIN**

Bi-stable images are not the unique type of optical illusions that shows convenient correspondences with quantum physics. There are some static optical illusions that are perceived to rotate, and can be very effective in explaining the concept of quantum spin.

Books and documentaries commonly describe the spin of a particle like the "rotation" of a marble around its axis, which can rotate clockwise or counterclockwise, but this is completely wrong. Elementary particles are point-like objects and, on the contrary of marbles, do not have a surface, nor a volume or any structure, so that the classical notion of rotation has no sense for them. Nonetheless, a particle with non-zero spin does have a mechanical angular momentum and, correspondingly, a magnetic moment if it is electrically charged, like e.g. an electron.

Moreover, an electron within an atom also possesses an orbital angular momentum and the corresponding magnetic moment, even if quantum mechanics clearly showed that electrons do not move around the nucleus as a planet does around the Sun.

*Quantum spin is in no way a classical angular momentum*, but it is *an angular momentum of a different form*. Like potential energy and thermal energy are very different forms of energy, but can be exchanged and produce the same effects, a particle with spin can transfer angular momentum to a macroscopic object and cause it to rotate. Quantum spin thus *exhibits the appearance of a rotation despite particles do not rotate in any way*.

I propose to illustrate this very counter-intuitive concept by means of "spinning illusion" drawings, like the ones in Fig 7. While looking at these figures the spectator perceives a clockwise or counterclockwise rotation even if the drawings do not rotate, in close correspondence with the notion of quantum spin that I outlined.

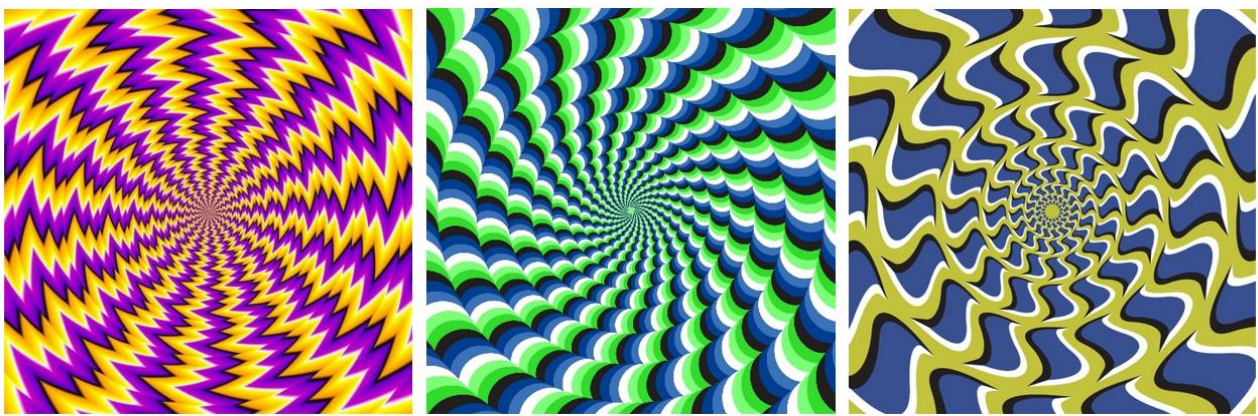

*Fig. 7: Spinning illusions*

The illusion of movement in such images generates from a suitable pattern of light and dark stripes by leveraging on the different response speed of retinal neurons to high contrast and low contrast colors. Variations of this pattern can lead to different directions of the apparent rotation (first panel in Fig 7 appears rotating clockwise, while the others seems counterclockwise), which can be used to represent spin up and spin down in a spin ½ particle.

To summarize:

- *quantum spin* → spinning illusion drawing.
- *spin up or down for a spin ½ particle* → clockwise or counterclockwise direction of rotation illusion.

The limitation of these drawings is that they cannot support spin superposition, since the same rotation is perceived when looking at them again.

**OVERPASSING THE LIMITATIONS**

As told, both the bi-stable and spinning illusions presented so far show some major limitations, namely i) the lack of support for the measurement process, ii) the inadequacy in representing entanglement when the two cards are distant from each other, or seen by different persons, and iii) the missing representation of spin superposition in the spinning illusions.

The world of visual ambiguities though is quite large, so that the teacher can adopt different illusions, getting different pros and cons. It is beyond the scope of this work to analyze the adequacy of any known illusion to be used for quantum representations, but it's worth to present an illusion able to support the measurement process, and another one representing the spin superposition.

Fig. 8 depicts the behavior of the "Close/Far" illusion, which belongs to the so-called "hybrid drawings". Looking at it the observer sees either one of the words "*CLOSE*" or "*FAR*" depending on its distance from the drawing.

This happens because the drawer predominantly used high spatial frequencies to compose the shape to be viewed nearby, and low spatial frequencies for the shape to be viewed by far. When the drawing is close, the attention is captured by the high-frequency figure, while when it is far the high frequencies vanished beyond the eye's resolution and the low-frequency figure prevails.

For each person there is a narrow range of intermediate distances where the hybrid drawing appears as bi-stable. To represent the *measurement of a quantum state in superposition*, the spectator shall place the drawing at such bi-stable distance, and suddenly move it closer or further away as soon as he reads either "*CLOSE*" or "*FAR*", respectively. Once moved, the drawing is no longer in a bi-stable condition and any further observation will confirm the previously "measured" state.

We could even use this scheme to represent the entanglement by placing two identical hybrid drawings one aside the other on a common support. As the spectator perceives *"CLOSE"* or *"FAR"* in one of the drawings, he correspondingly moves the support closer or further away carrying them out from the bi-stable region, so that the perceived shape will be viewed on both.

The same happens if we place the two drawings on a very wide support, so that they stay at a large separation from each other. This way a different person that will observe the second drawing after we have "measured" the first one, will see the same shape we randomly perceived on our side.

In comparison to a bi-stable drawing, this hybrid illusion has the advantage to deal with quantum measurement, but the disadvantage of being less practical and intuitive. So that they complement each other.

Figure 9 instead presents frames of the "Spinning ballerina" illusion, which is a bistable animation showing a dancer that can be seen spinning in one direction or in the other. This animation is able to represent the superposition of a one-half spin particle (if a video device is available), but it is not suitable to represent the "spin without spinning" concept. Therefore, this illusion complements those shown in Fig. 7.

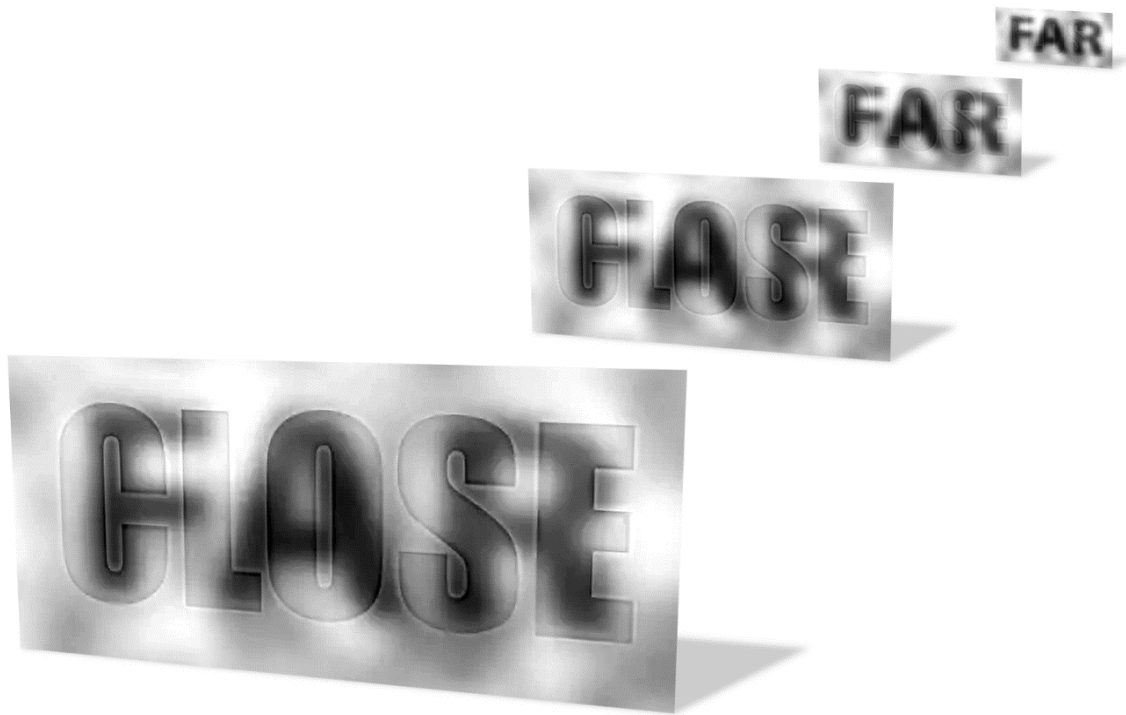

Fig. 8: "Close / Far" illusion

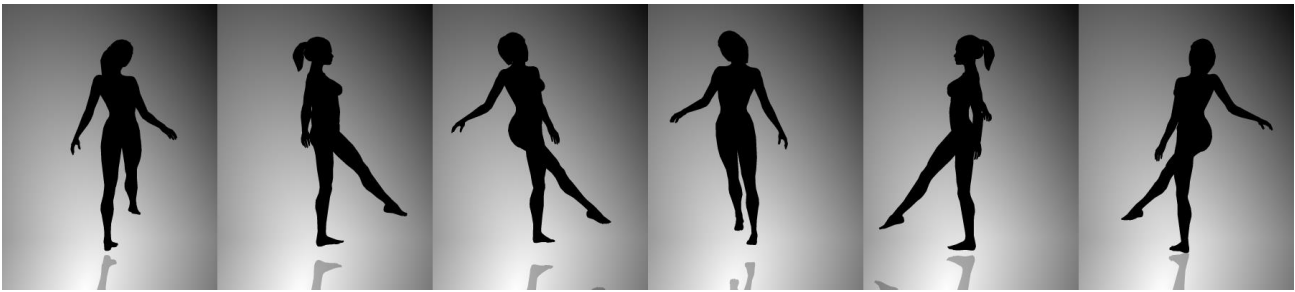

Fig. 9: The "Spinning ballerina" animation
(https://upload.wikimedia.org/wikipedia/commons/2/21/Spinning_Dancer.gif)

**CONCLUSIONS**

I present a new visual paradigm, which aims at exploiting optical illusions as an aid to explain the concepts of quantum mechanics.

In particular, I demonstrate the effectiveness of using a specific set of optical illusions, called "bi-stable drawings", to represent a quantum object. Thanks to their inherent duality, these drawings lead the spectator to understand the quantum concepts through a direct sensorial experience, which avoids most misconceptions.

Such analogy has a definite correspondence for a number of quantum concepts, like superposition, collapse, complementarity, and entanglement, which I discuss extensively together with their limits of applicability. Thanks to these correspondences, the spectator learns the meaning of state superposition, understands complementary of the states, and realizes that entangled particles does not inter-communicate and that quantum teleportation does not actually transport anything.

I also introduce the "spinning drawings" optical illusion, demonstrating its ability to represent the mysterious concept of quantum spin avoiding the common, but misleading, comparison to a classical rotation, and teaching to the spectator that nothing is actually spinning in quantum spin.

For each illusion, I analyze the limitations of the analogy, and finally I give two examples of how to overcome these limits by exploring the large word of visual ambiguities.

This work aims to invite the reader to exploit in lessons, presentations, and books the advantages of using inherently dual visual representations for explaining quantum concepts. Here I examined bi-stable, spinning, and hybrid optical illusions, but the reader could try exploring more sensorial ambiguities to represent the many counter-intuitive concepts of modern physics.

**AKNOWLEDGEMENTS**

The author likes to thank the team of the italian podcast of physics "Fisicast", and in particular professor Giovanni Organtini of La Sapienza University of Rome, for helpful discussions and comments.

**Gianluca Li Causi** is a senior researcher of the Rome Astronomical Observatory of the INAF, where he works in the field of astronomical technologies. Has worked for various instrument of different telescopes and is currently responsible for the data simulation of the next-generation multi-object spectrographs MOONS@VLT and MOSAIC@ELT, and for the data processing of the high-contrast camera SHARK-VIS@LBT, which has recently produced the highest-resolution ground-based image of the Jovian moon Io. He is active in outreach and didactics and has been co-founder and director of the Italian podcast for physics "FISICAST," for which he developed new ideas for explaining modern physics to the people.